\documentclass[a4paper,12pt,twoside,fleqn]{article}
\usepackage{graphicx}
\usepackage{times}

\newcommand{\sectionword}[1]{
  \begin{center}\stepcounter{section}\setcounter{subsection}{0}
    {\underline{\bf\normalsize \arabic{section}.~#1}}
  \end{center}
}
\newcommand{\subsectionword}[1]{
  \stepcounter{subsection}
    {\noindent\underline{\bf\normalsize
        \arabic{section}.\arabic{subsection}~#1}}\\ \indent
}

\newcommand{\nomenclature}[3]{$#1$ -- #2, #3\\}

\begin{document}
\setcounter{totalnumber}{20}
\setcounter{topnumber}{20}
\setcounter{bottomnumber}{20}
\setcounter{dbltopnumber}{20}
\title{\bf\large CAN ONE MAKE A POWDER FORGET ITS HISTORY?}
\author{{\bf\normalsize \underline{M. Morgeneyer$^1$},
    \underline{\vphantom{y}L. Brendel$^2$},} 
  {\bf\normalsize Z. Farkas$^2$, D. Kadau$^2$,}\\
  {\bf\normalsize D. E. Wolf$^2$, and J. Schwedes$^1$}\\
  {\normalsize$^1$Institute of Mechanical Process Engineering,}\\
  {\normalsize Technical University of
   Braunschweig, 38104 Braunschweig, Germany}\\
   {\normalsize$^2$Institute of Physics, University Duisburg-Essen,
   D-47048 Duisburg, Germany}}
%
%
%
\date{}

%
\maketitle
\begin{abstract}
It is shown that 
computer simulations can qualitatively reproduce experiments, where a powder of
cohesive, round, hard particles is periodically deformed at constant volume.
Two types of initial configurations are considered: Uniaxially
precompacted ballistic deposits and biaxially precompacted
DLA-clusters. Both initial configurations had the same volume
fraction, but due to the different precompaction procedure completely
different principal stresses.
After a transient which lasts only less than a period, the stresses
follow the same periodic function, i.e. the powder forgot its history.
\end{abstract}



\sectionword{INTRODUCTION}

One of the fundamental open questions in the physics of powders is 
how to characterize their state uniquely, the
particle configuration in general depending on the deformation history 
of the powder. This is in marked contrast to thermal equilibrium
systems, which statistically explore all available configurations,
so that temporal correlations normally die out quickly.
A powder, however, is ``jammed'' or frozen in, if it is not 
driven by external forces. Even if it flows, it does not explore 
its alternative configurations on a time scale which is short compared
to the inverse strain rate.

This has macroscopic consequences. For example the
stress in a compacted cohesive powder not only depends on the solid fraction
but also, by what kind of compaction the solid fraction was
reached. It is believed, that at least the fabric tensor is needed in
order to determine the stress state. 
Any microscopic theory of the dynamical
evolution of the state and hence of powder flow 
requires an answer to the question, how to characterize such a state.
Therefore it is very important also for macroscopic predictions needed
e.g. in process engineering.

In this paper we address only a partial aspect of this question.  It
is known, that the stress components in a powder, which is
quasi-statically deformed at fixed volume, approach values
independent of the precompaction history \cite{Nowak93}. This 
stress state
depends on the volume fraction, but also on the total deformation at
constant volume: As one continues to shear, the stress components increase.
The point we want to make
in this paper is that there are also other ways to make the powder
forget its precompaction history. As an example we investigate the
periodic kneading of the powder at constant volume. In contrast to
steady state flow the principal axes switch back and forth
periodically in this case. This has the advantage that 
the sample returns to its initial shape periodically.
In every period steady state flow properties may be
reached or not, depending on how long the period lasts.

Periodic kneading was studied in a different context previously
\cite{morgeneyer2002}. There it was shown for limestone that the stresses
are periodic functions with decreasing amplitude. After a few
periods the amplitude becomes constant. In the present paper we 
compare experiments on carbonyl-iron powder and computer simulations.
In contrast to limestone the particles are spherical and very hard.
As we are going to show, the amplitude of the periodic stresses
becomes constant already within one cycle in this case. 
Simultaneously all stress tensor differences due to different
sample preparations vanish within this short transient, i.e., the
powder forgets its history.

\sectionword{METHODS}

\begin{figure}
  \begin{center} 
    a) \includegraphics[width=0.4\linewidth]{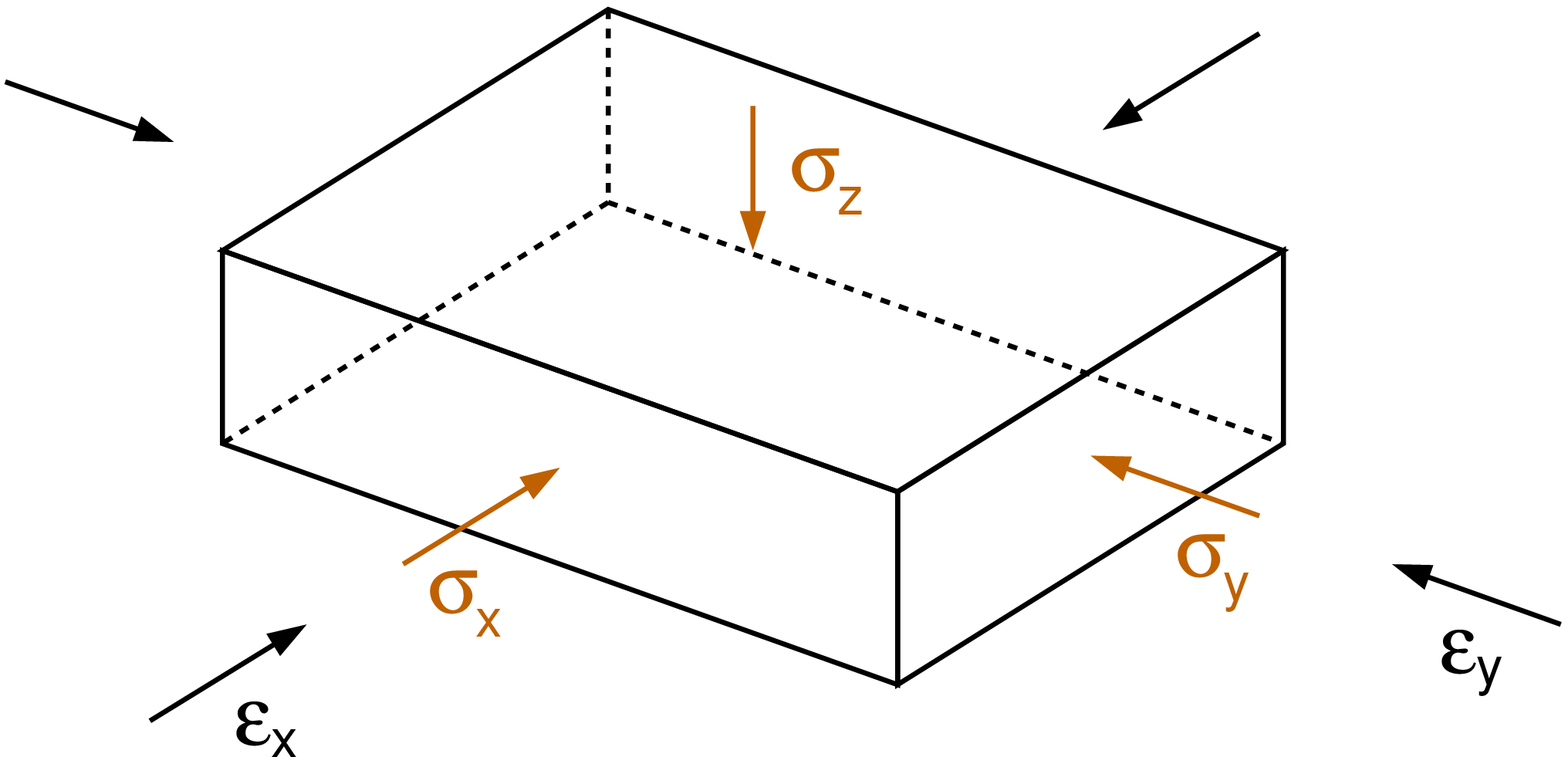} \hfill
    b) \includegraphics[width=0.45\linewidth]{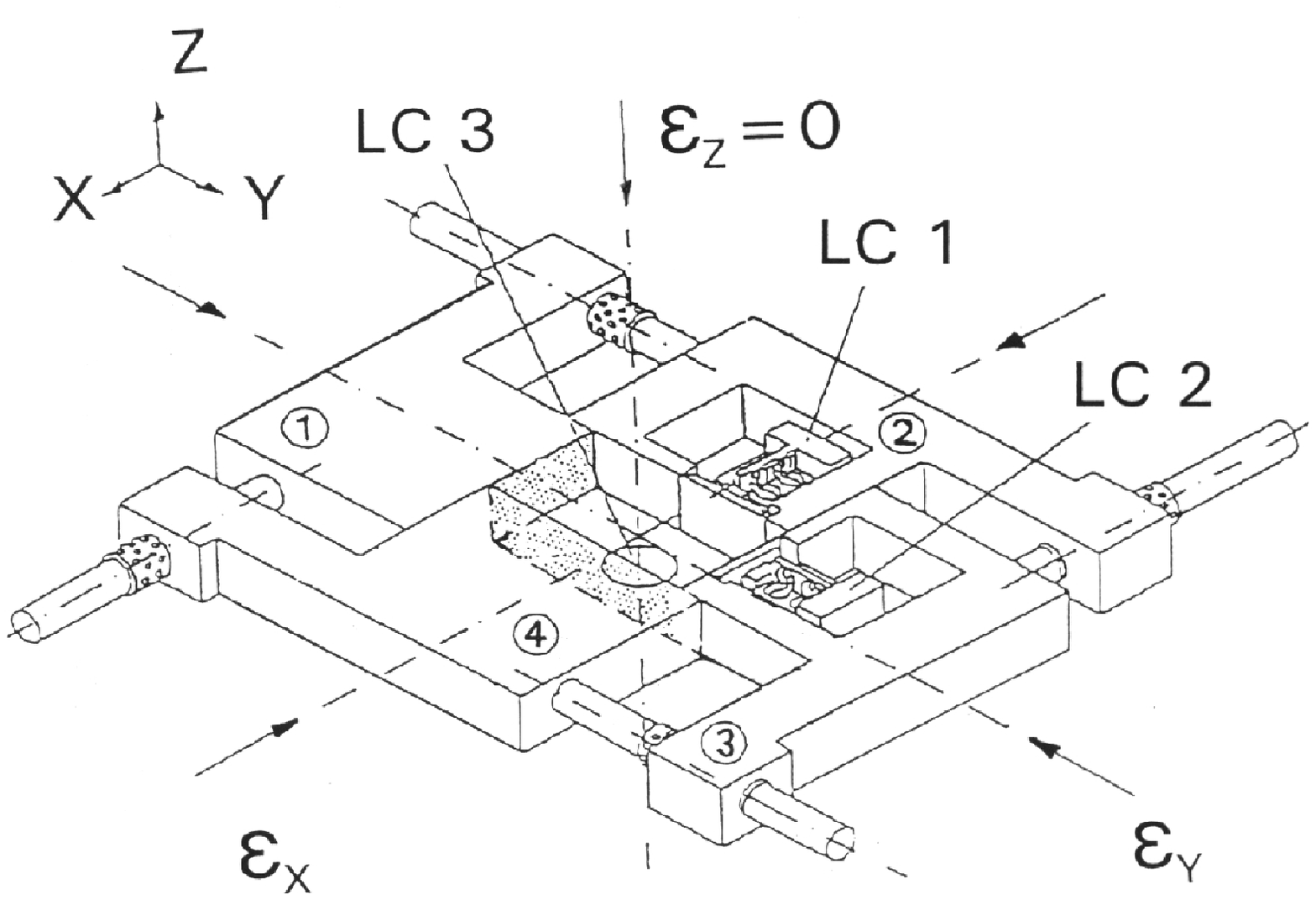}
  \end{center}
  \caption{ a) Schematic view of  the sample chamber explaining the
    axes.  b) The true biaxial shear tester.}
  \label{fig:boxandsample}       
\end{figure}

\subsectionword{Experimental Setup}
The true biaxial shear tester allows the deformation of a brick shaped 
bulk solid specimen. The deformation mechanism is composed of a bottom 
and a top plate which are fixed at a distance of 36.5 mm from each other 
and four side plates which can be moved independently from a maximum 
distance of 130 mm to a minimum distance of 70 mm, always staying 
perpendicular to their adjacent ones. Thus the bulk solid specimen 
remains brick shaped and principal strains equal the normal strains
(Fig.~\ref{fig:boxandsample}a): 
\begin{equation}
  \label{eq:strainprincipal}
\epsilon_{ij} = 0,\mbox{ where } i\not= j; \quad\epsilon_{ii} =
\epsilon_i \, .
\end{equation}
The side plate's velocities can be set in steps of 0.03 $\mu$m/s up to a 
maximum velocity
of 65 $\mu$m/s corresponding to maximum shear rates of 0.1\% per 
second. On the specimen's border, the complete stress state is measured 
with three five component load cells \cite{schulze89}, which are 
installed in the bottom and in two perpendicular load plates,
LC 1, 2, and 3 in Fig.~\ref{fig:boxandsample}b.
 For the tests with the true biaxial shear tester, silicon 
grease is spread on bottom, lateral and top plates. Then they are 
covered with highly flexible rubber membranes in such a way that the 
membranes are evenly deformed with the borders of the powder specimen. 
The friction between specimen and load plates thus is minimized, hence 
shear stresses are negligible. Therefore, the normal stresses on the 
specimen borders are principal stresses. The principal stress axes 
coincide with the deformation axes (Fig.~\ref{fig:boxandsample}a):
\begin{equation}
  \label{eq:stressprincipal}
\sigma_{ij} = 0,\mbox{ where } i \not= j;\quad \sigma_{ii} = \sigma_i.
\end{equation}
In this paper we use the engineering sign convention for the stress,
i.e. a compressive stress is defined as positive.
Deformations and stresses are continuously measured using a process 
computer, which also controls the deformation process
\cite{morgeneyer2002}.  As experiments with finite strains are carried out, we 
define the strains using the natural strain increment \cite{ludwik09} and
obtain
\begin{equation}
  \label{eq:ludwik}
  \epsilon_i =\ln \frac{L_i}{L_{i0}}
\end{equation}
for the strains in $x$-, $y$- and $z$-direction and for the volume strain 
respectively:
\begin{equation}
  \label{eq:ludwikvolume}
  \epsilon_V =\ln \frac{V}{V_0} =\ln
  \frac{L_xL_yL_z}{L_{x0}L_{y0}L_{z0}} =\epsilon_x + \epsilon_y +\epsilon_z
\end{equation}
In order to perform tests at constant volumes, one can easily control 
the condition $\epsilon_x + \epsilon_y =$const. ($\epsilon_z=0$ during
the complete measurement due to the fixed bottom and top plates). 
For test preparation, the powder is first sieved in $z$-direction into the 
tester. Then, it is precompacted in the same direction in order to 
homogenize the sample. After this, the sample is sheared with
$\dot\epsilon_x= -\dot\epsilon_y$, i.e., the 
sample's volume remaining constant. The directions of shear are changed 
periodically.
All experimental results presented in this paper have been obtained with 
carbonyl-iron powder with a median particle size of about 2 microns. 
This powder was chosen because its primary particles are perfectly
round, in analogy to the particles used in the computer simulations.

\subsectionword{Simulation Method}
For the computer simulations we use a discrete element method, the
Contact dynamics method \cite{jean99}. 
In contrast to soft particle molecular
dynamics, the volume exclusion of perfectly
rigid particles and Coulomb's friction law are implemented exactly
\cite{wolf96}.  The cohesion between the particles is included in a
simplified model as an attractive force acting within a range leading
to a force and an energy barrier to break a contact between two
particles \cite{kadau2002,kadau2003,radjai2001}. Furthermore, rolling
friction between the round particles is implemented, also leading to
the stabilization of large pores \cite{kadau2002,kadau2003}. We use
two dimensional simulation because the influence of the torsional
degree of freedom for a contact present in three dimensions is not
well understood up to now.


All simulation results are given in the following units: As unit of
length we take the particle radius $r$, as unit of mass 
$\rho r^3$, where $\rho$ is the mass density,
and as unit of time the inverse strain rate
$|\dot\epsilon_x|^{-1}=|\dot\epsilon_y|^{-1}$ (which was constant
during the periodic kneading). In these units the simulation
parameters were: The cohesion force between two particles, $F_{\rm c}
= 10^{22} \rho \, r^{4} \, |\dot\epsilon_x|^2$, the range of the
cohesion force, $d = 10^{-4}\, r$, the friction coefficient, $\mu =
0.3$, and the rolling friction coefficient, $\mu_{\rm R} = 0.2 \, r$ 
(which is the torque resisting the rolling motion divided by the
normal force $F_{\rm n} 
+ F_{\rm c}$ at the contact). We like to point
out that for the typical values of the carbonyl-iron powder experiment
[$r \approx 1$~$\mu$m, $\rho \approx 10^4$~kg/m$^3$ and
$|\dot\epsilon_x| \approx 10^{-4}$~s$^{-1}$] our simulation parameter
$F_{\rm c} \approx 10^{-6}$~N is a reasonable value for the cohesion
force between the particles.

The walls were cohesive, but frictionless, so that no tangential
stresses could occur, as in the experiment. This guarantees, that the
principal axes of the stress and the strain tensors were collinear.

\sectionword{RESULTS}

\begin{figure}[t!]
  \begin{center}
    \includegraphics[width=0.75\linewidth]{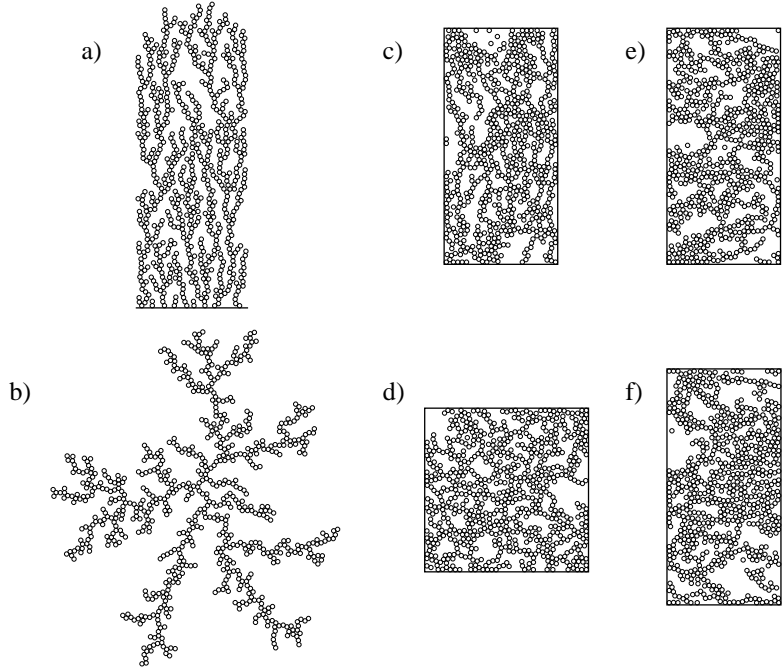} 
  \end{center}
  \caption{a) Initial configuration 1: Ballistic deposit. b) Initial
    configuration 2: DLA cluster. c) Configuration 1 precompacted
    uniaxially in the deposition direction. d) Configuration 2
    precompacted biaxially (keeping the square symmetry). e)
    Configuration 1 after one cycle. f) Configuration 2
    after one cycle.}
  \label{fig:configurations}
\end{figure}
For the two-dimensional simulations, we prepared two different kinds
of initial configurations in the following way (cf.\ 
Fig.~\ref{fig:configurations}): One is a ballistic deposit
\cite{meakin91}, which was precompacted in the direction of the
deposition \cite{kadau2003} up to a volume fraction of $\nu=0.42$.
The other is an off-lattice DLA cluster \cite{witten_sander81}
centered in a square box, and then precompacted biaxially to the same
volume fraction, shrinking
the square box symmetrically. Note that the first preparation
procedure tends to lead to an\-iso\-tro\-pic, but homogeneous
configurations, while the second one results in inhomogeneous, but
isotropic configurations (the initial DLA cluster is essentially round
and has a higher density at the center).  This was done deliberately
to demonstrate that even in this case both configurations reach the
same (periodic) stress/strain relationship.

\begin{figure}
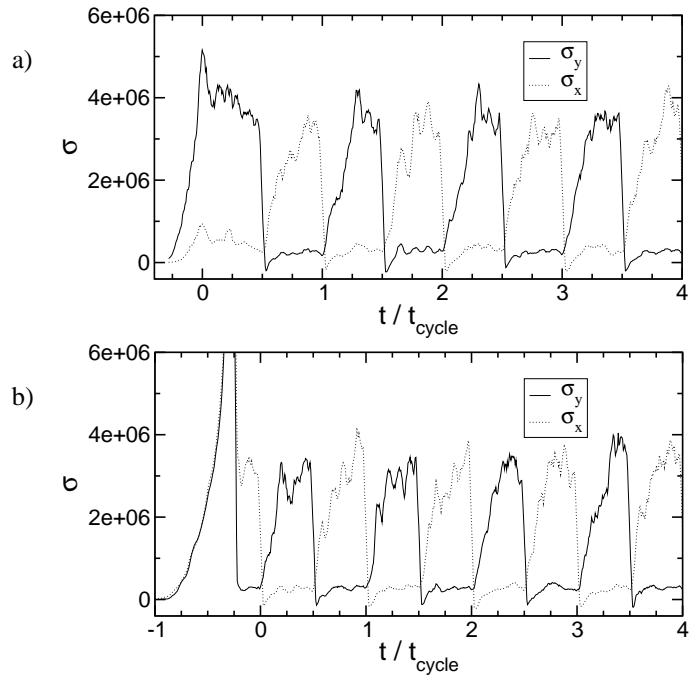

  \begin{center}
    \includegraphics[scale=0.325,angle=0]{ballistic42_smooth.eps}\hbox 
    to 1cm{}
    \includegraphics[scale=0.325,angle=0]{dla42_smooth.eps}\hbox
    to 1cm{}
  \end{center}
  \caption{Simulation results: Principal stresses as functions of
    time during periodic deformation a) for configuration type 1
    [volume constant for $t / t_{\rm cycle} \geq 0$], and b) for
    configuration type 2 [volume constant for $t / t_{\rm cycle} \geq
    -1/4$].  }
  \label{fig:stresses_sim}
\end{figure}
\begin{figure}
  \begin{center}
    \includegraphics[scale=0.35, angle=0]{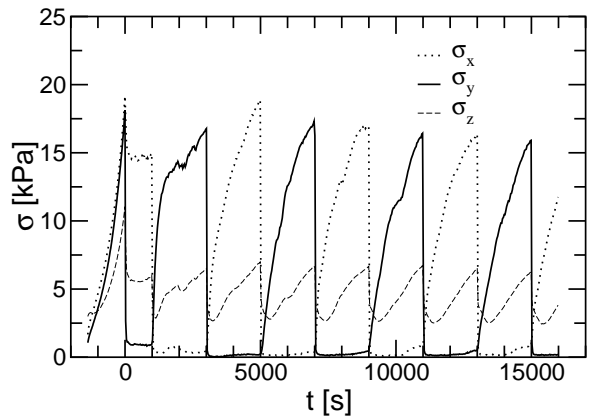}    
  \end{center}
  \caption{Experimental results: Principal stresses as functions of
    time during periodic deformation of biaxially precompacted
    carbonyl-iron powder. 
    }
  \label{fig:stresses_exp}
\end{figure}
The precompacted systems are now used as initial configurations for
the periodic shearing at constant volume. As in the experiment, the
shearing was performed with constant logarithmic strain rate. The
maximal strain was $36\%$ (with respect to the square shape). Figure
\ref{fig:stresses_sim} shows the time dependence of the two principal
stresses, averaged over $20$ independent runs (with different, but
similarly prepared initial configurations), including the
precompaction. Moreover the data were smoothed by averaging them over
a running time window of 5\% of the period. 


Clearly, the stresses are qualitatively different during the
precompaction: For configuration type 1 (ballistic deposit) and
uniaxial precompaction one gets an\-iso\-tro\-pic stresses, the larger
stress belonging to the direction of compaction. For configuration
type 2 (DLA-cluster) and biaxial precompaction both principal stresses
increase in the same way, but when the square is deformed into a
rectangular shape at constant volume (first 1/4 cycle), the principal
stress in the expanding direction drops to a small positive value,
while the other component stays large. This behavior agrees
qualitatively with the experiment with the biaxial precompaction,
Fig.~\ref{fig:stresses_exp}.

The time dependent stresses within the first half cycle ($t/t_{\rm
  cycle}=0$ to $1/2$) in Fig.~\ref{fig:stresses_sim}a are
qualitatively different from the ones in all subsequent half cycles.
Similarly the first 1/4 cycle ($t/t_{\rm cycle}=-1/4$ to $0$) in
Fig.~\ref{fig:stresses_sim}b is different.  However, after these
transients, it seems that the same stress pattern is repeated period
after period, also in the experiment Fig. \ref{fig:stresses_exp}. In
order to compare the stress patterns for the two cases, Fig.
\ref{fig:stresses_sim}a and b, we averaged them over 9 repetitions.
\begin{figure}[b]
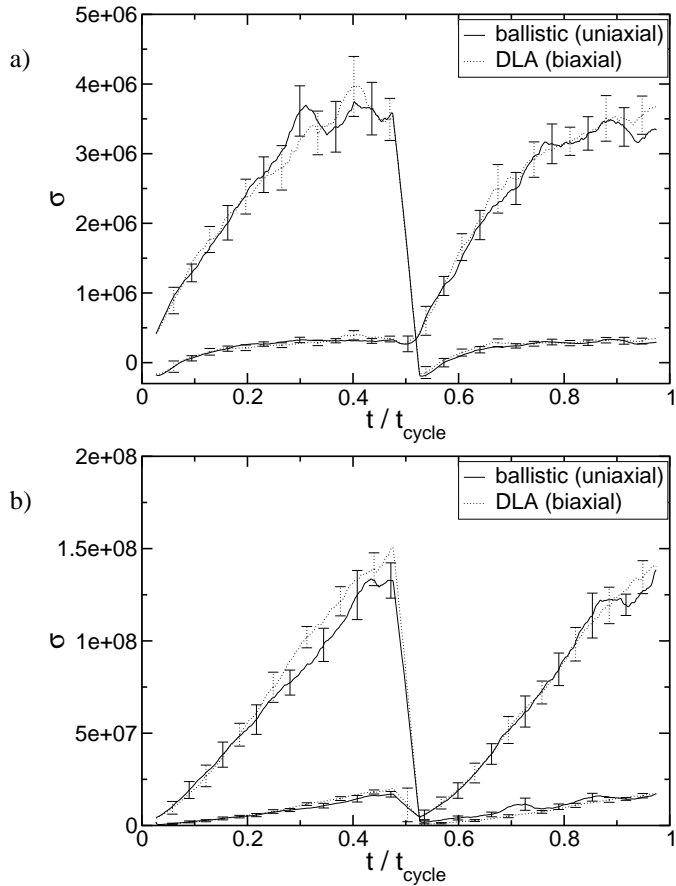

  \begin{center}
    \includegraphics[scale=0.325]{period-avg42_1graph_smooth.eps}
    \includegraphics[scale=0.325]{period-avg63_1graph_smooth.eps}
  \end{center}
  \caption{One period of $\sigma_x$ and of $\sigma_y$, averaged over
    all repetitions 
    after the initial transient. a) For $\nu = 0.42$ as in
    Fig.~\ref{fig:stresses_sim}, b) for $\nu = 0.63$. In both cases
    the curves for initial configurations of type 1 [full line] respectively
    type 2 [dashed line] coincide within the statistical errors.
    }
  \label{fig:stress_comparison}
\end{figure}
The results are shown in Fig.\ref{fig:stress_comparison}. Within the
error bars the stress pattern for the samples obtained from uniaxially
precompacted ballistic deposits is the same as for those obtained from
biaxially precompacted DLA-clusters. Moreover $\sigma_y$ simply
coincides with $\sigma_x$ shifted by half a period, as it should. This
shows, that after the transients the stresses do not show any
dependence on the preparation history. This is also reflected in
Fig.\ref{fig:configurations}e and f, which show that after one cycle
the particle configurations are qualitatively indistinguishable.

Figure \ref{fig:stress_comparison}b shows the stress pattern
for a higher solid fraction. Here the maximum strain compared to the
square shape was 16\%. Again one gets the same pattern in spite
of the different types of original configurations and  precompaction
histories. However, the shape and amplitude of this pattern is
completely different. This shows that the periodic stress pattern 
depends on the solid fraction.

\sectionword{CONCLUSIONS}

We investigated the strain-dependent stress  in a cohesive powder, which 
was periodically deformed at constant volume and strain rate.
It was shown that 
contact dynamics simulations agree qualitatively with experiments on
carbonyl-iron powder.

For the simulations two types of initial configurations 
were prepared. One had an\-iso\-tro\-pic stress and fabric tensors
(uniaxially compacted ballistic deposits), the other was isotropic
(biaxially compacted DLA-clusters), but both types of configurations had
the same volume fraction. These initial configurations were then 
periodically deformed at constant volume. 
After a transient which lasts only less than a period,
one obtains for both types of initial configurations
the same periodic stress pattern. This leads us to the conclusion that
periodic shear like
steady state flow makes the powder forget its history quickly.
It was also shown, that the periodic stress pattern depends on the solid
fraction.

\sectionword{REFERENCES}\vspace{-2cm}
\def\refname{}
\bibliographystyle{unsrt}
\bibliography{gran_mat,nano,friction,sonstige}
%

\sectionword{NOMENCLATURE}

\par\noindent
\nomenclature{\epsilon}{strain tensor}{dimensionless}
\nomenclature{\sigma}{stress tensor}{Pa}
\nomenclature{L}{size of the sample chamber}{m}
\nomenclature{V}{volume of the sample chamber}{m$^3$}
\nomenclature{r}{particle radius}{m}
\nomenclature{\rho}{mass density}{kg m$^{-3}$}
\nomenclature{d}{range of the cohesion force}{m}
\nomenclature{F_c}{cohesion force}{N}
\nomenclature{F_n}{normal force}{N}
\nomenclature{\mu}{coefficient of tangential friction}{dimensionless}
\nomenclature{\mu_{\rm R}}{coefficient of rolling friction}{m}
\nomenclature{\nu}{volume fraction}{dimensionless}

\sectionword{ACKNOWLEDGMENT}

We thank Guido Bartels and Henning Knudsen for fruitful discussions.
This research was supported by DFG within grants WO 577/3-1, SCHW
233/29-1 and SFB 445, as well as by Federal Mogul GmbH and BASF AG
Ludwigshafen.

\end{document}